\documentclass[twocolumn]{aastex631}

\bibpunct{(}{)}{;}{a}{}{,}
\usepackage{graphicx}
\usepackage{amsmath}
\usepackage{gensymb}
\usepackage{cleveref}
\usepackage{tabularx}
\usepackage{booktabs}
\usepackage{comment}
\usepackage{rotating}
\usepackage{txfonts}
\usepackage{hyperref}
\hypersetup{
    colorlinks=true,
    linkcolor=blue,
    filecolor=magenta,      
    urlcolor=blue,
    citecolor=blue
}

\accepted{ApJ}
\shorttitle{The High-redshift Blazar: \target~}
\shortauthors{Sanchez Zaballa, J.M., Bottacini, E., Tramacere, A.}
\usepackage{array}
\usepackage{changepage}
\usepackage{ulem}

\RequirePackage{soul}
\RequirePackage{xcolor}

\newcommand{\target}{MG3 J163554$+$3629}
\newcommand{\T}{$\rm{T1}~\left(\rm{1}\right)$}

\begin{document}

\title{The High-redshift Blazar \object{MG3 J163554+3629}: Physical Properties and the Enigma of Its Unexpected Supermassive Black Hole Growth}

\author{Jose Maria Sanchez Zaballa}
\affiliation{Dipartimento di Fisica e Astronomia ``G. Galilei'', Università di Padova, I-35131 Padova, Italy}
\affiliation{Lehrstuhl f{\"u}r Astronomie, Universit{\"a}t W{\"u}rzburg, Emil-Fischer-Strasse 31, W{\"u}rzburg, 97074, Germany}

\author[0000-0001-9579-0487]{Eugenio Bottacini}
\affiliation{Dipartimento di Fisica e Astronomia ``G. Galilei'', Università di Padova, I-35131 Padova, Italy}
\affiliation{W.W. Hansen Experimental Physics Laboratory \& Kavli Institute for Particle Astrophysics and Cosmology, Stanford University, USA}
\affiliation{Eureka Scientific, 2452 Delmer Street Suite 100, Oakland, CA 94602-3017, USA}
\affiliation{INFN Sezione di Padova, Via Marzolo 8, I-35131 Padova, Italy}

\author[0000-0002-8186-3793]{Andrea Tramacere}
\affiliation{Department of Astronomy, University of Geneva, Ch. d’Ecogia 16, 1290 Versoix, Switzerland}

\correspondingauthor{Eugenio Bottacini}
\email{ilbotta4@gmail.com}

\begin{abstract}

There is general consensus that active galactic nuclei (AGNs) derive their radiating power from a supermassive black hole (SMBH) that accretes matter.
Yet, their precise powering mechanisms and the resulting growth of the SMBH are poorly understood, especially for AGNs at high redshift. Blazars are
AGNs pointing their jet toward the observer, thus being detectable from radio through gamma rays at high redshift due to Doppler boosting.
The blazar \target~is located at redshift $z=3.65$ and it is a flat spectrum radio quasar (FSRQ). In this work, we show the results of the modeling of its
spectral energy distribution (SED) from radio to gamma rays with a one-zone leptonic
model. We estimate the uncertainties through a Markov Chain Monte Carlo approach. As a result, we infer the black hole mass
$M_{BH}=1.1^{+0.2}_{-0.1}\times10^{9} M_{\odot}$ and a modest magnetic field of $B=6.56^{+0.13}_{-0.09}\times10^{-2}~G$ in line with the Compton dominance
observed in high-redshift FSRQs. The emitting region is outside the broad line region but within the region of the dust torus radius. The rather
small accretion efficiency of $\eta=0.083$ is not solely inferred through the SED modeling but also through the energetics. An evolution study suggests that
in an Eddington-limited accretion process the SMBH did not have time enough to grow from an initial seed mass of $\sim10^{6} M_{\odot}$ at $z\approx30$
into a mass of $M_{BH}\approx10^{9} M_{\odot}$ at $z=3.65$.  Faster mass growth might be obtained in a super-Eddington process throughout frequent
episodes. Alternative scenarios propose that the existence of the jet itself can facilitate a more rapid growth.

\end{abstract}


\section{Introduction} \label{sec:intro}
Galaxies harbor at their center supermassive black holes (SMBHs) as an inevitable consequence of galaxy and stellar evolution.
The speed at which these SMHBs grow is an interesting topic, even more so for those at high redshift with masses of the order
of $M$~$\sim$~10$^{7-10}$~\(M_\odot\). These black holes had little time (on a cosmological scales) to grow to such high masses
and two leading scenarios are proposed that invoke the merger between two or more black holes and the accretion at a high rate onto
the black hole  \citep[e.g.][]{yoo04, tanaka09}.
These SMBHs descend from seed black holes formed through stars or gas in the first
galaxies \citep{rees78}. At $\sim$ $10^{3} M_{\odot}$, these seed black holes are divided into light seeds, which stem from the evolution of the first stars
with initial masses of $10^{2-3} M_{\odot}$ \citep[e.g.][]{natarajan14} and heavy seeds, which originate from the direct collapse of
primordial gas clouds with initial masses of $10^{4-6} M_{\odot}$ \citep[e.g.][]{volonteri08}. Observationally, seed black holes remain
unconstrained.\\
However, the detection and characterization of high-redshift (z$>$3) SMBHs and their environment (within a few parsec from the 
black hole including the accretion disk, the broad-line region, and the dust torus (DT)) is not straightforward due to their distance from
the observer.
A geometric arrangement that facilitates the detection and study of active galactic nuclei \citep[AGNs;][]{urry95} occurs when SMBHs form
jets that point in the direction of the observer, which are called blazars.
Historically, blazars are classified into two main types: flat-spectrum radio quasars (FSRQs) and BL Lac objects (BL Lacs). FSRQs are
characterized by broad, strong emission lines, whereas BL Lacs typically exhibit weak emission lines, if any \citep{urry95}. BL Lacs display
a larger amplitude of X-ray flux change compared to FSRQs \citep{sambruna97}. However, the primary distinction between these two classes is
typically made based on the equivalent width of the emission lines, using a threshold of 5 \text{\AA}. Yet, this classification can be somewhat
ambiguous, as both the emission lines and underlying continuum radiation can vary over time, complicating
the definition.
Blazars are bright on the entire electromagnetic spectrum 
from radio frequencies to gamma-ray energies \citep{urry95}. This allows for a rather precise modeling of their spectral energy distribution (SED)
including the modeling of the SMBH it grew into and the environment it is embedded in. The detection of high-redshift blazars at gamma-ray
energies \citep[mostly by the {\em Fermi} mission; ][]{Ajello22} is due to their high jet-power output \citep{celotti08}.
Therefore, gamma-ray data are rather constraining in modeling the SED and in understanding the physics at work that generates the jet power.
One of the preferred mechanism that generate astrophysical jets (on stellar and on galaxy scales) is the so-called Blandford-Znajek process
\citep{blandford77}, which requires the SMBH to be rotating. Such rotating black holes have been observed in AGNs by analyzing X-ray
spectral signatures from the accretion disk resulting from relativistic effects given the gravity in the vicinity of
the SMBH \citep[e.g.][]{reynolds19, bambi21}. Indeed, the radius of the innermost stable circular orbit (ISCO) of the accretion disk
depends on the dimensionless spin parameters of the black hole $|a|$~$<$1 \citep{bardeen73}: the more rapidly the SMBH spins,
the closer the accretion disk can reach to the SMBH thereby being subject to stronger relativistic effects. The jet originating from the
center of the galaxies powers continuously double radio sources \citep{rees71}. The rapid radio variability suggested relativistic motion and
beaming effects due to small observing angles, all of which was modeled through the dynamics of particle acceleration and
magnetic fields providing a continuous energy supply from the central environment close to the SMBH \citep{scheuer74, blandford74}. 
More precisely, general relativistic magnetohydrodynamic simulations of jets, which account for the self-consistent interaction between the
accretion disk and black hole, show that rotating black holes can generate jets, with jet power increasing as the black hole spin increases
\citep{mckinney06}.
Thus, the jet power depends on the accretion rate \citep{ghisellini14}, which on long timescales has been linked to galaxy-galaxy mergers \citep{padovani16}.
This would allow for explaining the decrease over cosmic time of powerful jets detected at gamma rays above 15 keV \citep{ghisellini10, volonteri11}
because of the decrease over time of the rate at which galaxies merge.
Therefore, the observation and the modeling of the 
SED emitted by blazar jets at high redshift represent rather rare opportunities to study their nuclear environment and their evolution.\\
The SED of blazar jets is usually modeled through two broad-band nonthermal emission components: synchrotron radiation at low energies
(between radio and X-rays) and inverse-Compton radiation at high  energies (above X-rays). Also thermal components from the accretion disk,
the broad-line region, and the dust torus can be modeled.
The peak-frequency of the synchrotron radiation spans a wide range from radio wavelengths to X-rays. According to its position
in this range blazars are divided into low- and high-synchrotron peaked for increasing frequencies \citep{padovani95}.
Very specifically, high-redshift blazars are Compton dominated \citep{celotti08}. This means that the luminosity radiated at high energies due to
inverse-Compton dominates over the luminosity at low energies due to synchrotron. Also, given that the latter peaks at low frequencies
this allows for the thermal radiation from the accretion disk to surface above the nonthermal radiation as shown in multiwavelegths studies \citep[e.g.][]{bottacini10}.
In turn, this spectral feature allows for estimating the accretion rate and the black hole mass. At the transition between
the synchrotron radiation and the inverse-Compton radiation a large fraction of the emitted radiation from the high-redshift blazar can be 
efficiently absorbed by neutral hydrogen along the line of sight. These photons have a rest-wavelength blue-wards of 912 \(\text{\AA}\),
which is called the Lyman limit, and 1216 \(\text{\AA}\) that is the Lyman-$\alpha$ forest. 
The resulting strength of the break due to the absorption can be computed through SED fitting. Since the result depends on the distance between the observer and the source, it can be used as a tool to estimate the redshift \citep[e.g.][]{rau12}.
As a consequence
a blazar will go undetected when observed with filters having responses blue-wards of the redshifted Lyman limit of (1+z)~$\times$~912 
\(\text{\AA}\).\par
A further redshift-dependent attenuation of the emitted radiation from blazars is given by its annihilation with the extragalactic background light (EBL) photon field. In more detail, the EBL is the result of the integrated emission of all stars and accreting compact objects in the observable Universe. This includes also those objects during the reionization epoch. The intensity of the EBL is rather uncertain because it is much fainter compared to the intensity of the bright foreground radiation due to the zodiacal light and the Milky Way \citep[e.g.][]{matsuoka11}. Despite its faintness the EBL imprints its distinct signature onto the blazar photon spectrum, while these photons propagate from the cosmological distance of the blazar to the observer. Very specifically the rest-frame GeV and TeV blazar photons annihilate with the EBL photons on the background light in the UV, optical, and IR bands \citep[e.g.][]{stecker92}. Therefore, the farther the blazar, the more its spectrum will be affected by attenuation. \par
Over the last decades, the development of modeling techniques and the advancements in observational capabilities (from space and the ground) across the electromagnetic spectrum have established the multiwavelength SED modeling of blazars as a standard tool. This approach is particularly valuable for constraining physical quantities in the centers of galaxies, especially in the context of AGN research.
This is particularly interesting for sources at high redshift, as they present a significant challenge in constraining physical properties due
to their remoteness. Importantly, their remote location provides a unique opportunity to study the evolution of SMBHs, as these objects have had
less time to grow compared to their lower-redshift counterparts.
\\
In this research we present the results of the modeling of the average SED of \target~in the leptonic particle scenario. This blazar is at redshift z=3.65 \citep{alberati17}.
Its excellent spectral coverage is a key asset in modeling the SED. In particular, the detection by the {\em Fermi} Large Area Telescope
spans a unique energy range, which is crucial for constraining the properties of the circum-nuclear environment.
Its position (R.A.=248.946816$^{\circ}$, Dec=36.491658$^{\circ}$) is found through precise astrometry through the VLBA imaging at 5 GHz \citep{petrov11}. \target~ is classified as a flat spectrum radio quasar \citep[FSRQ;][]{abdollahi22}.\par
Throughout this research we assume a flat $\Lambda$CDM cosmology with $\rm{H}_{\rm 0} = 67.8~\rm{km}~\rm{s}^{-1}~\rm{Mpc}^{-1}$, $\Omega_{\rm 0} = 0.307$ and $\Omega_{\Lambda} = 0.693$.

\section{Observations and data analyses of \target} \label{sec:observations}
In the following, we provide a description of the data from radio frequencies to gamma-ray energies. As a brief summary, the time intervals during which each instrument collected multiwavelength
data are as follows: radio data from 1990 to 2007, {\em WISE} data from 2009 to 2011, {\em GAIA} data from 2013 to 2020, {\em Swift}/UVOT data in 2016, {\em XMM}-Newton data in 2017, and {\em Fermi}
LAT data from 2008 to 2022. \Cref{tab:sed}, at the end of this section, reports the actual data.
\subsection{NRAO Green Bank Telescope \& Very Large Array} \label{subsec:NVSS}
We use the data available from different surveys: The FIRST Survey: Faint Images of the Radio Sky at Twenty Centimeters \citep{becker95}, The NRAO VLA Sky Survey \citep{condon98}, A New Catalog of 53,522 4.85 GHz Sources \citep{becker91}, The Third MIT--Green Bank 5 GHz Survey \citep{griffith90}, The 87 GB Catalog of Radio Sources Covering 0$^\circ$ $<$ $\delta$ $<$ +75$^\circ$ at 4.85 GHz \citep{gregory91}, A New Catalog of 30,239 1.4 GHz Sources \citep{white92}, A New Catalog of 53,522 4.85 GHz Sources \citep{becker91}, The VLBA Imaging and Polarimetry Survey at 5 GHz \citep{helmboldt07}. We used these data to locate the radio counterpart of the blazar and to build the SED, including two upper limits in the transition region between radio and IR frequencies in the observed frame. \par

\subsection{Wide-field Infrared Survey Explorer}\label{subsec:WISE}
The Wide-field Infrared Survey Explorer \citep[WISE; ][]{wright10} is part of NASA's Explorers Program \footnote{\url{https://explorers.gsfc.nasa.gov}} which was put into orbit in 2009, and it is one of the most suitable instruments to gather data in the IR band. Two years later after its launching, by 2011, it had surveyed the sky twice with its 40~cm telescope in four infrared wavelengths $w$1,$w$2, $w$3, and $w$4 at 3.4, 4.7, 12, and 23 $\mu$m, respectively. \target~ is detected by the telescope onboard the \textit{WISE} mission at a significance above 5$\sigma$ in the AB system in the two bands $w$1=18.91 ($\pm$ 0.05) mag and  $w$2=19.42  ($\pm$ 0.13) mag, while for the two bands $w$3 and $w$4, only upper limits at a 3~$\sigma$ level are derived. Conversions from instrumental source magnitudes to calibrated magnitudes were performed using the \textit{WISE} photometric systems \citep{wright10, jarrett11}.
Early studies of {\em WISE} data have shown the connection between IR waveband and gamma-ray emission from blazars \citep{plotkin12}. Such a connection has been explored in a sophisticated study by \cite{massaro11} and \cite{massaro16} who explain a physical origin in favor of their IR -- gamma-ray connection between the {\em Fermi}--LAT detected blazars and the {\em WISE} counterparts, where the latter often is given by the thermal emission.\par

\subsection{Gaia}\label{subsec:GAIA}
The {\em Gaia} mission \citep{prusti16} was launched in 2013 aiming to perform the most precise mapping of the Milky Way. While {\em Gaia} is optimized for selecting and measuring parallaxes and proper motion of point-like sources, it also provides accurate multi-band photometry. In a precise astrometry study by \cite{plavin19} related to the disk-jet environment in AGNs, the authors show that the coordinates and fluxes of the \textit{Gaia} counterparts of high-redshift blazars are driven by the contribution of the accretion disk over the synchrotron component from the jet. Therefore, also the photometry is dominated by the accretion disk confirming several multifrequency campaigns of high-redshift blazars \citep[e.g.][]{bottacini10}.\par
The \textit{GAIA} counterpart of the \textit{Fermi}--LAT source (4FGL J1635.6+3628) has been searched in the publicly available third data release \citep[DR3; ][]{vallenari23} of the {\textit Gaia} mission. \textit{GAIA}'s counterpart in GAIA DR3 of the {\textit Fermi}--LAT blazar is DR3 1327964378519702912. The {\em Gaia} G magnitudes are $G=20.7951\pm 0.0248$ mag, $G_{\text{BP}}=20.8728 \pm 0.3497$ mag, and $G_{\text{RP}}=19.6487 \pm 0.2105$ mag that are in agreement with the association and values of a cosmological evolution study of  \textit{Fermi}--LAT blazars by \textcite{zeng21} for which the authors used the {\em Gaia} DR2 \citep{brown18}.

\subsection{Swift UV--Optical Telescope}\label{subsec:Swift}
The high-redshift blazar has been targeted by the UV–Optical Telescope \citep[UVOT;][]{roming05} on board the {\em Swift} satellite between 2016-09-15 00:01:15 UTC and 2016-09-15 04:51:56 UTC (obs id: 00034715001) and more recently between 2023-10-27 07:26:52 UTC and 2023-10-27 07:26:52 UTC (obs id: 00089697001). These observations used the filters $V$, $U$, and $B$. The data are analyzed with HEASoft version 6.30 using CALDB version 20211108. This version allows for checking for the detector small-scale sensitivity, which can be corrected for count rates of up to 35\%. We used the standard 5$'$ aperture for the source and a much larger area of 50$''$ to adequately sample the background. This is used by the standard pipeline to reduce the image products and correct the exposure within the XIMAGE environment. The images of multiple observations have been summed for each filter, which in turn allows for computing the source fluxes. The on-source exposure for the $V$, $U$, and $B$ filter are 1.52466~ks, 1.10625~ks, and 1.165679~ks, respectively. 
The source is detected with the $B$ filter and marginally with the $U$ filter only. The resulting magnitudes have been been converted to flux density units
(1.3$\pm$0.1)$\times$10$^{-29}$ erg s$^{-1}$ cm$^{-2}$ Hz$^{-1}$ 
and (1.7$\pm$0.9)$\times$10$^{-30}$ erg s$^{-1}$ cm$^{-2}$ Hz$^{-1}$,
for the $B$ and the $U$ filter respectively) using the constants of the photometric system in \cite{poole08}.\par
\subsection{XMM-Newton}\label{subsec:XMM}
ESA's X-ray mission \textit{XMM}-Newton \citep{jansen01} has observed the source starting in 2017-07-23 10:27:12 (OBS-ID: 0802000101) for a rather low exposure (for such a distant source) of 23 ksec. Data are analyzed using the Science Analysis Software (\texttt{SAS}) 15.0.0 package following the standard procedure. The spectrum is extracted from a source region of $40''$ radius centered at the source of interest. The background is extracted from a larger area of $60''$ radius from the same chip, yet without contamination from the source itself or any other source. We use the tool \texttt{evselect} for extracting the source and background spectra. The tools \texttt{rmfgen} and \texttt{arfgen} are used to generate response files and ancillary files. The spectrum is well fit with a simple power-law model with neutral hydrogen absorption fixed to the Galactic value, which is N$^{gal}_{H}$ = 1.8$\times$10$^{20}$ atoms cm$^{2}$ obtained from the LAB Survey of Galactic HI database \citep{kalberla05}. \par

\subsection{Fermi Large Area Telescope}\label{subsec:FERMI}
The {\em Fermi} mission was launched in 2008. It observes the gamma-ray sky at GeV energies between 0.1 GeV and 300 GeV with its primary instrument the Large Area Telescope \citep[LAT; ][]{atwood09}.
For the detection of gamma rays in a photon-by-photon mode, the LAT uses a high-resolution silicon tracker, a CsI calorimeter, and it rejects the charged particles with an anticoincidence detector. Due to its observing strategy and its large field of view ($\sim$2.4 sr),
the LAT scans the entire sky every two orbits ($\sim$3 hours). This makes the LAT
very suitable for survey studies. For the gamma-ray range of the SED we use the LAT data available from the fourth \textit{Fermi}-LAT
source catalog - Data Release 4 \citep[4FGL-DR4; ][]{ballet23}. 
The blazar \target, with its 4FGL designation 4FGL J1635.6+3628, is detected at $\sim$ 11$\sigma$ and identified in
the LAT survey, from which the spectrum is taken.
The spectrum is best fit with a simple power-law model having a rather steep spectral
index of $\Gamma$=2.67 and a flux of
$F_{\gamma}$=(1.40$\pm$0.05)$\times$10$^{-12}$ erg cm$^{-2}$ s$^{-1}$.
\par
\begin{table}[htbp]
\centering
\begin{tabular}{lcl}
\hline
\hline 
Observation 	& Frequencies		&  Fluxes \\
			& [Hz]			&  [erg~cm$^{-2}$ s$^{-1}$]	\\
\hline 
\\			
NRAO VLA (radio)		& 1.40$\times$10$^{9}$	&(1.55$\pm$0.07)$\times$10$^{-15}$ \\
NRAO VLA$^{*}$ (radio)	&1.40$\times$10$^{9}$	&1.58$\times$10$^{-15}$ \\
NRAO GB	$^{*}$ (radio)	&1.40$\times$10$^{9}$	&1.72$\times$10$^{-15}$ \\
NRAO VLA (radio)		&1.40$\times$10$^{9}$	&(2.13$\pm$0.06)$\times$10$^{-15}$ \\
NRAO GB	$^{*}$ (radio)	&4.85$\times$10$^{9}$	&2.66$\times$10$^{-15}$ \\
NRAO GB	 (radio)		&4.85$\times$10$^{9}$	&(4.17$\pm$0.62)$\times$10$^{-15}$ \\
NRAO GB	 (radio)		&4.85$\times$10$^{9}$	&(4.32$\pm$0.58)$\times$10$^{-15}$ \\
NRAO VLBA$^{*}$ (radio)	&5.00$\times$10$^{9}$	&3.04$\times$10$^{-15}$ \\
WISE$^{*}$ (IR)		&1.32$\times$10$^{13}$	&2.94$\times$10$^{-13}$ \\
WISE$^{*}$ (IR)		&2.42$\times$10$^{13}$	&7.22$\times$10$^{-14}$ \\
WISE (IR)				&6.30$\times$10$^{13}$	&(3.89$\pm$0.44)$\times$10$^{-14}$ \\
WISE (IR)				&8.53$\times$10$^{13}$	&(8.43$\pm$0.40)$\times$10$^{-14}$ \\
GAIA (optical)			&3.25$\times$10$^{14}$	&(4.84$\pm$0.99)$\times$10$^{-14}$ \\
GAIA (optical)			&3.89$\times$10$^{14}$	&(7.35$\pm$0.29)$\times$10$^{-14}$ \\
GAIA (optical)			&4.77$\times$10$^{14}$	&(6.87$\pm$0.26)$\times$10$^{-14}$ \\
Swift-UVOT (optical)		&6.17$\times$10$^{14}$	&(3.91$\pm$0.24)$\times$10$^{-14}$ \\
Swift-UVOT (optical)		&8.36$\times$10$^{14}$	&(4.00$\pm$2.23)$\times$10$^{-15}$ \\
XMM-Newton (X)		&1.63$\times$10$^{17}$--1.46$\times$10$^{18}$	&(4.90$\pm$0.93)$\times$10$^{-14}$ \\
Fermi-LAT (gamma)		&1.71$\times$10$^{22}$--7.65$\times$10$^{25}$ 	&(1.40$\pm$0.05)$\times$10$^{-12}$ \\ \bottomrule
\end{tabular}
\caption{The table reports the observations and their fluxes as used for the SED. $^{*}$observations represent values of upper limits.} 
\label{tab:sed}
\end{table}

\section{SED and Its modeling} \label{sec:SED}
The line of sight to \target~ is little affected by extinction. Nevertheless, we correct for this attenuation. We infer the extinction with Cardelli's law 
\citep{cardelli89} with a reddening coefficient of R$_{V}$=3.1 and an extinction coefficient of A$_{V}$=0.055  \citep{schlegel98}.\\
The SED of \target~exhibits the typical two broad and nonthermal radiation components that are characteristic of blazars \citep[e.g.][]{fossati98}. The low-energy component is
understood in terms of synchrotron radiation by relativistic electrons in the jet. The high-energy component is due to inverse-Compton scattering on 
photon fields that originate internally and/or externally to the jet by the same relativistic electrons. Its peak position can be hinted by the {\em Fermi}-LAT
spectrum. The onset of the high-energy component is located between the {\em XMM}-Newton spectrum and the optical data point of {\em Swift}-UVOT's
$B$ filter. The optical flux at wavelengths shorter than $\sim$5000 \textup{~\AA} ($\sim$5.9$\times$10$^{14} Hz$) is significantly suppressed by the absorption due to the intervening
Ly $\alpha$ \citep{weymann81}.  This explains why the source goes undetected at shorter wavelength in the $U$ band and it also explains the sharp
increase of the SED towards lower frequencies.
The non-detection by {\em Swift}-UVOT with the $V$ filter can be due to its in-orbit effective area, which is only $\sim$50\% of that of the $B$ filter (in which
the source is detected) as shown in the detailed research in \cite{poole08}, which in turn limits the sensitivity of the observation. The SED displays in its
low-energy component the signature of the bright thermal radiation from the accretion disk typical for high-redshift blazars due to their large inverse-Compton to 
synchrotron ratio \cite{ghisellini09}.\par
To model the overall SED, we adopt the one-zone model, for which the leptonic processes take place in a volume of spherical geometry of radius $R$.
This emitting volume is at a distance $R_{diss}$ from the SMBH, whose mass is $M_{BH}$. The spherical (in the comoving frame) emission region is
observed within an angle $\theta$ and it moves with a bulk Lorentz factor $\Gamma$ (corresponding to a speed of $\beta$$c$) in the jet. This leads to
a Doppler factor $\delta$~=~($\Gamma$~[1-~$\beta$~cos~$\theta$])$^{-1}$. The radiating volume is permeated by a randomly oriented magnetic field
having strength $B$. The very same volume is also subject to injection of relativistic leptons whose energy distribution is a broken power-law model 
described by:
\begin{gather}
\begin{aligned}
    n(\gamma) &= Kf(\gamma)\\
    f(\gamma)&=
    \begin{cases}
      \gamma^{-p}, & \gamma_\text{min}\leq \gamma<\gamma_\text{break} \\
      \gamma^{-p_{1}}, & \gamma_\text{break}\le \gamma\le \gamma_\text{max}
    \end{cases}
\end{aligned}
\end{gather}
where $p$ and $p_{1}$ are the low and high spectral indices respectively. Additionally, $\gamma_{min}$, $\gamma_{break}$, and $\gamma_{max}$ are
the minimum, break, and maximum electron Lorentz factors, respectively. $K$ is the normalization constant.

The number of emitting particles per unit volume is given by:
\begin{equation}
\mathrm{N} = \int_{\gamma_{min}}^{\gamma_{max}} n(\gamma)\:d\gamma .
\label{eq-crab}
\end{equation} 

In this context \texttt{JetSeT} \citep{tramacere09, tramacere11, tramacere20} is able to simulate and reproduce the
emerging radiation and acceleration processes, in which the injected relativistic electrons hit the magnetic field giving off synchrotron radiation
and inverse-Compton (IC) radiation on low-energy photons. These photons can originate external to the jet \citep[EC e.g.,][]{marscher85, bloom96} from the accretion disk \citep[e.g.,][]{dermer92}, from a fraction of the disk's
luminosity reflected off the Broad Line Region (BLR) \citep[e.g.,][]{sikora94} or the dusty torus (DT) in the infrared 
\citep[e.g., ][]{ghisellini96, bazejowski00}, or from synchrotron radiation between decelerating and accelerating relativistic
flows \cite[e.g.,][]{georganopoulos03}. Alternatively, the photons can originate within the jet itself from the synchrotron radiation,
in which case the IC radiation is synchrotron self-Compton (SSC). Following \cite{kaspi07} and \cite{cleary07}, the BLR is 
assumed to have a radius of $R_{BLR}$=3.00$\times$10$^{17}$ ($L_{disk,46}$)$^{0.5}$ cm, while the radius of the DT is estimated to be
$R_{DT}$=2.00$\times$10$^{19}$ ($L_{disk,46}$)$^{0.5}$ cm \citep{bazejowski00}. Both regions reprocess isotropically 10\%
($\tau_{BLR}$=0.1 ,$\tau_{DT}$=0.1) of the disk luminosity $L_{disk}$ in the rest frame of the SMBH.\\
The surfacing of the thermal emission feature of the accretion disk allows for confidently estimating the physical parameters of the accretion
disk. Its luminosity is inferred through the actual optical and IR data.
We use them to fit a multi-temperature (10$^{3}$ - 10$^{4}$ K) black body.
The bolometric flux $F_{bol}$ is computed using a Monte Carlo integration method, which allows in turn to calculate the disk luminosity through
$L_{disk}$~=~4$\pi$$D^{2}_{L}$$F_{bol}$~=~(1.381$\pm$0.003)$\times$$10^{46}$ erg s$^{-1}$.
Also, being $\eta$ the accretion efficiency and $\dot{M}$ the mass accretion rate, these parameters are related through
$L_{disk}$~=~$\eta$$\dot{M}$$c^{2}$. The initial estimate of the accretion disk luminosity allows for a handle on the initial fit conditions that
affect the size of the SMBH and the IC components at high energies. We account for the attenuation of the high-energy flux due to the annihilation
with the EBL photons using the model provided in \cite{franceschini08}.\\
The jet can carry its power in different forms. These forms are the radiation power ($P_{rad}$), the power carried by the magnetic field ($P_{B}$,
also known as Poynting flux), the kinetic power carried by electrons ($P_{e}$), and the kinetic power carried by cold protons ($P_{p}$, assuming
one proton per emitting electron). As a result of the SED modeling, we can compute these powers using the following equation:\\ \\ 
$P_{i}$=$\pi R^{2} \Gamma^{2} \beta c U'_{i}$\\ \\ 
where $R$ is the radius of the emission volume and $U'_{i}$ refers to the energy density measured in the comoving frame of the $i$-th form.
Very specifically, $P_{rad}$ incorporates the single radiation components from synchrotron, SSC, and the scattered radiation on the BLR and the
DT. The total jet power $P_{jet}$ is given by the sum of these single components.

\begin{table}[htbp]
\centering
\begin{tabular}{l|cl}
\toprule
\textbf{Parameter} & \textbf{Value} & \textbf{Units}\\ \midrule
$R$                               		& $0.79$  &  $10^{17}$ cm            \\
$R_{\text{diss}}$           		& $1.50$  &  $10^{18}$ cm            \\
$B$                               		& $6.56$  &  $10^{-2}$ G             \\
$\theta$                        		& $3.00$  &  deg                    \\
$\Gamma$                   		& $1.90$  &  $10^{1}$                \\
$\delta$                         		& $1.91$  &  $10^{1}$                \\
$\gamma_{\text{min}}$		& $1.00$  & \nodata                        \\
$\gamma_{\text{break}}$           & $0.94$  &  $10^{3}$                \\
$\gamma_{\text{max}}$             & $2.00$  &  $10^{4}$                \\
$N$                               		& $9.21$  &  $10^{2}$ cm$^{-3}$      \\
$p$                               		& $1.45$  &  \nodata                       \\
$p_{1}$                           		& $3.58$  &  \nodata                  \\
$R_{\text{in}}$                   		& $9.27$  &  $R_{\text{S}}$           \\
$R_{\text{out}}$                  	& $8.41$  &  $10^{2} R_{\text{S}}$  \\
$M_{BH}$                         		& $1.10$  &  $10^{9} M_{\odot}$      \\
$\eta$                            		& $0.83$  &  $10^{-1}$               \\
$L_\text{disk}$                   	& $1.41$  &  $10^{46}$ erg s$^{-1}$ \\
$R_{\text{DT}}$                   	& $2.37$  &  $10^{19}$ cm           \\
$T_{\text{DT}}$                   	& $9.27$  &  $10^{2}$ K             \\
$\tau_{\text{DT}}$                	& $0.10$  &   \nodata                  \\
$R_{\text{BLR}_{\text{in}}}$       & $3.56$  &  $10^{17}$ cm            \\
$R_{\text{BLR}_{\text{out}}}$	& $3.92$  &  $10^{17}$ cm            \\
$\tau_{\text{BLR}}$              	& $0.10$  &   \nodata                      \\ \bottomrule
\end{tabular}
\caption{SED parameters of the best-fit model, which refers to \Cref{fig:PrePremodel}. $R_{\text{S}}$ stands for Schwarzschild radius} 
\label{tab:PrePre_model}
\end{table}

\section{Results and Discussion}
\subsection{SED Fit Results}
\begin{figure*}[ht!]
\centering
    \includegraphics[width=\textwidth]{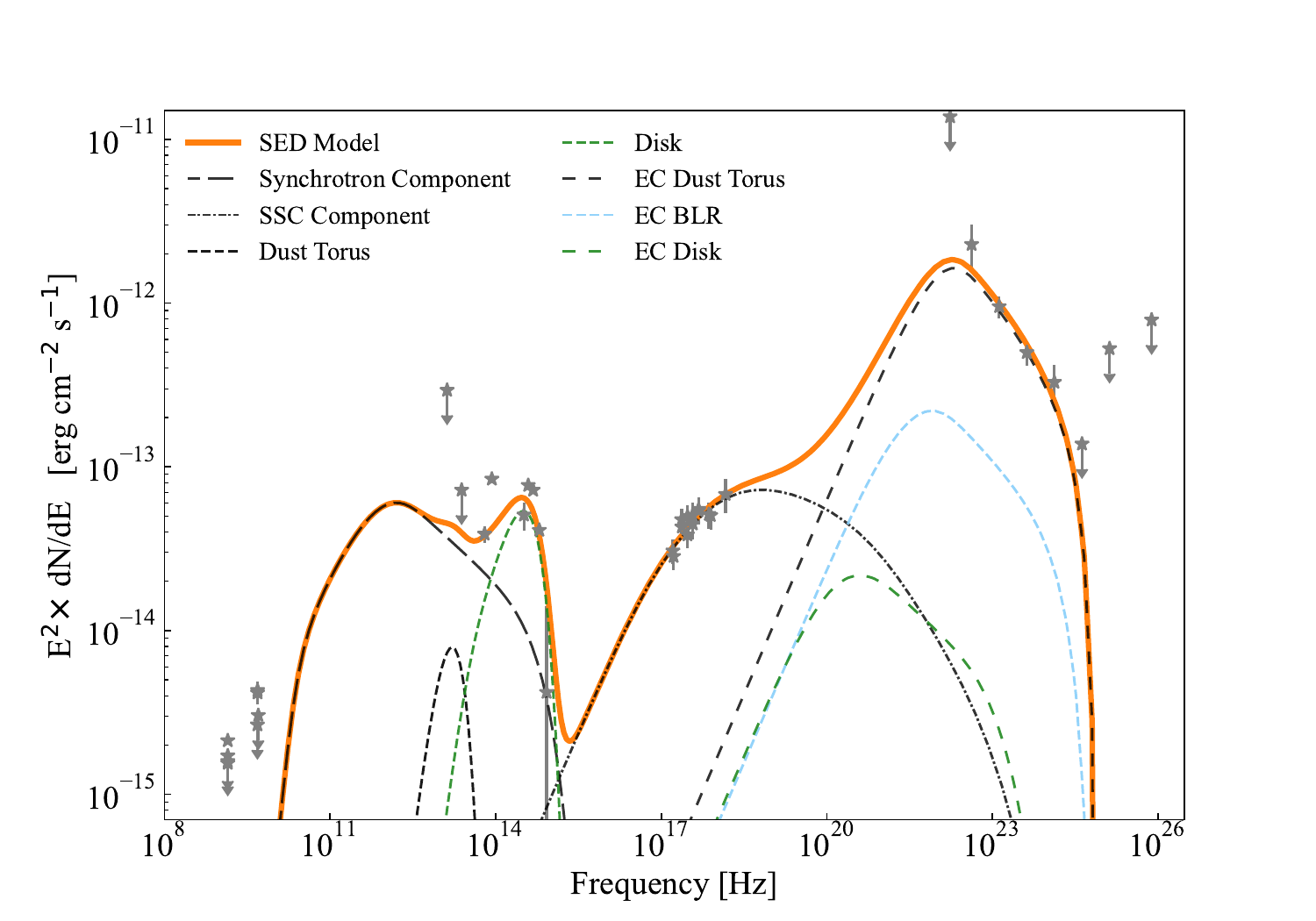}
    \caption{Observed-frame SED data and best-fit model along with the single emission components for \target~. In \Cref{tab:PrePre_model} we report the corresponding
    parameters including $M_{BH}$=1.1$\times$10$^{9}$ $M_{\odot}$ and the accretion disk efficiency $\eta$=0.083. The dissipating region is beyond the BLR but within the radius
    of the DT.}
    \label{fig:PrePremodel}
\end{figure*}

The estimate of the thermal emission from the accretion disk as described above allows for a rather good initial SED fitting values. Also, it provides
an excellent estimate for the high-energy component due to the seed photons from the disk. In addition, the good quality of the data by the LAT and
{\em XMM}-Newton allow to constrain the SSC and EC components. The former component largely depends on the energy distribution of the injected particles,
on the magnetic field (and thus is linked to the synchrotron component), and on the the bulk Lorentz factor. The EC component has a dependency
on the location of the emission within the emitting volume. Since we have a good initial estimate of the thermal emission from the accretion disk
and its contribution to the EC, also the location of the emitting volume can be confidently inferred. Given its resulting distance from the SMBH of
$R_{diss}$=1.5$\times$10$^{18}$ cm, which is outside the BLR (where the outer radius is
$R$$_{BLR}$$_{out}$=3.92$\times$10$^{17}$cm) but 
within the radius of the DT $R_{DT}$=2.37$\times$10$^{19}$ cm, the EC component from the DT is expected to dominate over the component
from the BLR. The magnetic field is rather modest of $B$=0.06 Gauss and the bulk Lorentz factor of $\Gamma$=19 is a typical value for FSRQs \citep{ghisellini14}.
These latter two values are in  good agreement with theoretical predictions for FSRQs in a recent work by \cite{rueda21}, which
also predicts a concurrent steep gamma-ray spectrum (based on the injected particle-energy distribution) that is confirmed
by \target's power-law photon spectral index of 2.81. Studies of time lags between gamma-ray and radio emission of large samples \citep[e.g.][]{pushkarev10}
suggest that their emission regions are different, where the radio emission possibly originates in external layers of the jets and from a different electron
population \citep{finke16}. Such different locations of the emission region are a challenge to properly reproduce the radio data with the models as shown in
many studies \citep{tramacere22}. We find a black hole mass of $M_{BH}$=1.1$\times$10$^{9}$ $M_{\odot}$. This value is in good agreement with $M_{BH}=10^{8.7} M_{\odot}$
inferred though IR-optical data only \citep{ackermann17}. From the overall SED modeling the accretion disk luminosity is
$L_{disk}$=1.41$\times$$10^{46}$ erg s$^{-1}$. The inner and the outer radii of the accretion disk result in
$R_{in}$=9.27 $R_{S}$ and  $R_{out}$=8.41$\times$10$^{2}$ $R_{S}$, where the unit $R_{S}$ is the Schwarzschild radius
($R_{S}$=2$G$$M_{BH}$/$c^{2}$). The accretion efficiency is $\eta$=0.083. The SED parameters are reported in \Cref{tab:PrePre_model}, while 
\Cref{fig:PrePremodel} shows the results in the $E^{2}$~$\times$~$dN/dE$ representation.

The same figure shows that the X-ray data constrain the SSC component. The major contributor to the inverse-Compton component at its peak (EC DT) is the
dust torus. A somewhat smaller contribution (EC BLR) is provided by the BLR being beyond the location of the emitting volume. The comptonized accretion
disk radiation (EC Disk) is given by the bright accretion disk, which is fit through the optical data. The data point from the {\em Swift}-UVOT $B$ filter falls short
from being properly fit because of the attenuation of these photons due to the absorption by neutral hydrogen (Lyman limit) along the line of sight between
the source at cosmological distance and the observer. \\

\subsection{Accretion Efficiency}
\begin{figure*}[ht!]
\centering
    \includegraphics[width=\textwidth]{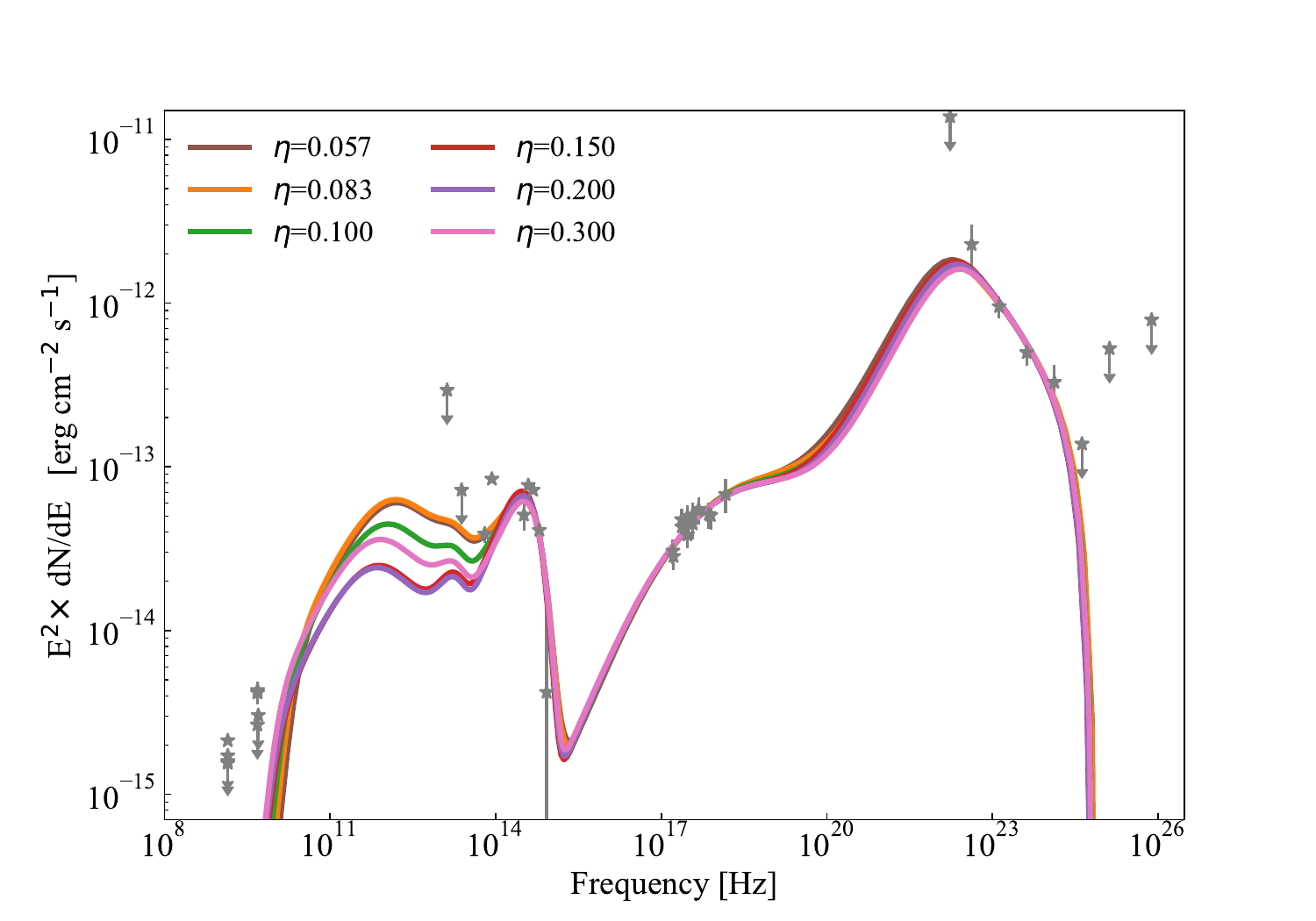}
    \caption{Observer-frame SED models for different accretion efficiencies, $\eta$=0.083 being the best-fit result. Higher accretion efficiencies (and thus
    higher disk luminosities) impact the BLR radii increasing their sizes. For every accretion efficiency, the dissipating region is beyond the BLR but within the radius
    of the DT.}
    \label{fig:etas}
\end{figure*}

Building upon the dependence of the accretion disk luminosity by the accretion efficiency $\eta$, we use this latter variable
to explore how it affects the modeling of the physical parameters of the SED.  We analyze arbitrary but quasi homogeneously spaced accretion
efficiencies between $\eta$=0.057 for a  Schwarzschild black hole (no spin $a$=0) and $\eta$=0.300 for a Kerr black hole
(maximum spin $a$$\sim$0.998) \citep{thorne74}.
We show the inferred SED parameters in \Cref{tab:etas}. In \Cref{fig:etas} we show the different SEDs for each $\eta$. 
The most notable differences
among the models occur at frequencies below $\sim$10$^{13}$ Hz as a result of different locations of the emitting 
regions and sizes of the dust torus.
For the parameters that could be affected most by the accretion
efficiency we estimated their uncertainties with a Markov Chain Monte Carlo (MCMC) approach using a maximum likelihood test. Following the formalism
by \cite{sawicki12}, we defined a likelihood function representing the probability of the dataset given the model parameters. The
likelihood function is expressed as a sum of squared deviations between the data and model, weighted by the data errors. The errors are
assumed to be Gaussian and independent. By maximizing the likelihood function, we minimize the sum of squared deviations, which leads to
the formulation of the $\chi^2$ statistic. Incorporating upper limits in the error function, we construct the posterior probability function using
Bayesian data analysis. A flat prior is employed maintaining a constant probability centered on the best-fit values for each parameter.
The MCMC analysis involves a burn-in run consisting of 10 steps to approach equilibrium, followed by a production run of 75 steps with 128 walkers.
The results are summarized in \Cref{tab:MCMC}
showcasing the model parameters within the 2$\sigma$ confidence intervals of the SED. As a result all the parameters have well defined
uncertainties.\\
The accretion disk efficiency, and thus the disk luminosity, determines how much radiation is available to impact the size and luminosity of the BLR.
Therefore, the two quantities should be physically related. In fact, for increasing efficiencies also the radii of the BLR increase, which is supported by an
observational relation in a sample of blazars \citep[e.g.][]{kaspi05}. Also, for small efficiencies the dissipating region is well beyond the BLR, yet
well within the radius of the DT.\\

\begin{table}[htbp]
\centering
\footnotesize
\begin{tabular}{l|lllllll}
\toprule
\textbf{Parameter}            & \multicolumn{6}{c}{\textbf{Value}}                            & \textbf{Units}          \\ \midrule
$\eta$                  			& 0.057 & 0.083 & 0.100 & 0.150 & 0.200 & 0.300 & \nodata \\ \midrule
$R$                           			& $0.63$  & $0.79$ & $0.94$ & $1.10$ & $1.22$ & $1.42$ & $10^{17}$ cm            \\
$R_{\text{diss}}$             		& $1.20$  & $1.50$ & $1.80$ & $2.10$ & $2.33$ & $2.70$ & $10^{18}$ cm            \\
$B$                           			& $7.95$  & $6.56$ & $4.05$ & $2.16$ & $1.90$ & $2.23$ & $10^{-2}$ G             \\
$\theta$                      		& $3.00$  & $3.00$ & $3.00$ & $3.00$ & $3.00$ & $3.00$ & deg                     \\
$\Gamma$                      		& $2.00$  & $1.90$ & $1.57$ & $1.32$ & $1.32$ & $1.52$ & $10^{1}$                \\
$\delta$                      		& $1.91$  & $1.91$ & $1.87$ & $1.78$ & $1.79$ & $1.86$ & $10^{1}$                \\
$\gamma_{\text{min}}$         	& $1.00$  & $1.00$ & $1.00$ & $1.00$ & $1.00$ & $1.00$ & \nodata                        \\
$\gamma_{\text{break}}$      	& $0.88$  & $0.94$ & $1.07$ & $1.24$ & $1.28$ & $1.23$ & $10^{3}$                \\
$\gamma_{\text{max}}$        	& $2.00$  & $2.00$ & $2.00$ & $2.00$ & $2.00$ & $2.00$ & $10^{4}$                \\
$N$                           			& $11.48$ & $9.21$ & $7.70$ & $9.01$ & $8.17$ & $6.03$ & $10^{2}$ cm$^{-3}$      \\
$p$                           			& $1.42$  & $1.45$ & $1.41$ & $1.41$ & $1.41$ & $1.45$ & \nodata                        \\
$p_{1}$                       		& $3.61$  & $3.58$ & $3.65$ & $3.68$ & $3.66$ & $3.63$ & \nodata                       \\
$R_{\text{in}}$               		& $9.27$  & $9.27$ & $9.27$ & $9.27$ & $9.27$ & $9.27$ & $R_{\text{S}}$   \\
$R_{\text{out}}$              		& $8.41$  & $8.41$ & $8.41$ & $8.41$ & $8.41$ & $8.41$ & $10^{2} R_{\text{S}}$   \\
$M_{BH}$                      		& $1.15$  & $1.10$ & $1.24$ & $1.46$ & $1.30$ & $1.15$ & $10^{9} M_{\odot}$      \\
$L_\text{disk}$               		& $0.99$  & $1.41$ & $1.99$ & $3.37$ & $4.22$ & $5.78$ & $10^{46}$ erg  s$^{-1}$ \\
$R_{\text{DT}}$              		& $1.99$  & $2.37$ & $2.82$ & $3.67$ & $4.11$ & $4.81$ & $10^{19}$ cm            \\
$T_{\text{DT}}$               		& $9.27$  & $9.27$ & $9.27$ & $9.27$ & $9.27$ & $9.27$ & $10^{2}$ K              \\
$\tau_{\text{DT}}$           		& $0.10$  & $0.10$ & $0.10$ & $0.10$ & $0.10$ & $0.10$ &  \nodata                       \\
$R_{\text{BLR}_{\text{in}}}$  	& $2.98$  & $3.56$ & $4.23$ & $5.51$ & $6.17$ & $7.21$ & $10^{17}$ cm            \\
$R_{\text{BLR}_{\text{out}}}$ 	& $3.28$  & $3.92$ & $4.66$ & $6.06$ & $6.78$ & $7.93$ & $10^{17}$ cm            \\
$\tau_{\text{BLR}}$           		& $0.10$  & $0.10$ & $0.10$ & $0.10$ & $0.10$ & $0.10$ &  \nodata                      \\ \bottomrule
\end{tabular}
\caption{SED models for the set of fixed accretion efficiencies $\eta$ of \target~. The corresponding models are shown in \Cref{fig:etas}.} 
\label{tab:etas}
\end{table}

\begin{figure}[htbp]
\includegraphics[width=0.47\textwidth]{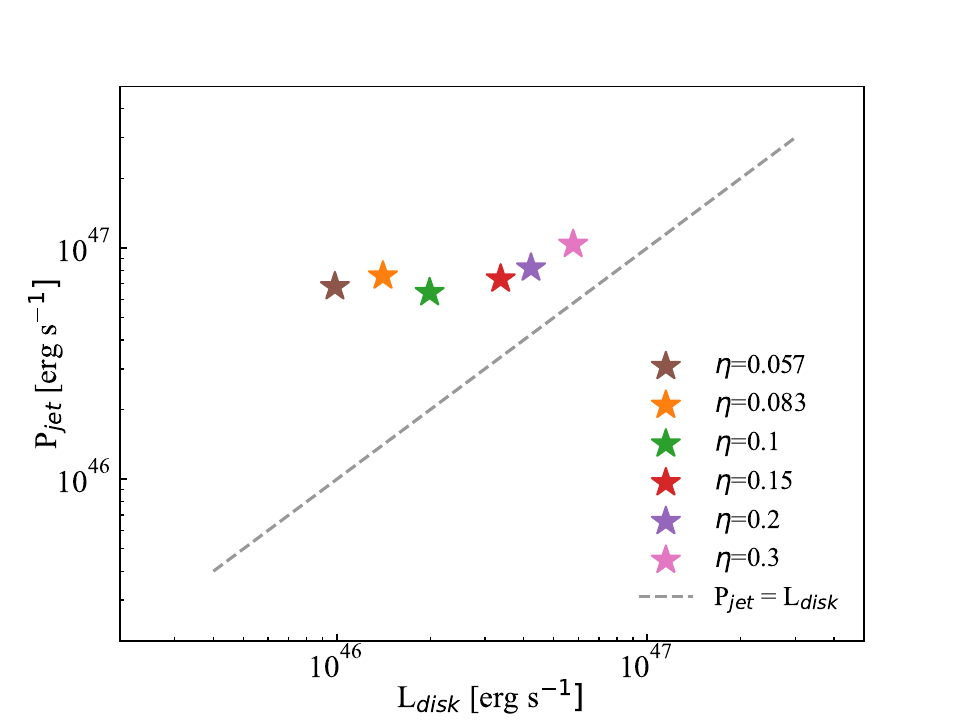}
\caption{P$_{jet}$ as function of L$_{disk}$ for the set of efficiencies $\eta$. The dashed gray line represents the equation P$_{jet}$=L$_{disk}$. Accretion
disk efficiencies of $\eta$$<$0.15 are favored over higher values (see text for discussion) thereby supporting the SED best-fit model.}
\label{fig:pvl}
\end{figure}

The spin of the SMBH, its mass, and the jet power are intrinsically related through the accretion process. 
For the different efficiencies we compute the power carried by the jet in different forms as defined above. They are reported in Table \ref{tab:jetpower}.
For the lowest three efficiencies $\eta<0.15$, the total kinetic power associated to the bulk motion of electrons and protons ($P_{e}$+$P_{p}$) dominates
over the radiated power $P_{rad}$ by a factor of 34 to 52. This result agrees with the findings in a population study of blazars for which the factor
 is shown to be within 10 and 50 \citep{celotti08, maraschi03}. Very similar results are found in studies of smaller samples of blazars
\citep[e.g][]{sambruna06}. These large total jet powers provide the necessary energy supply for the leptons to emit up to gamma-ray energies.
The contribution of the Poynting flux $P_{B}$ to the total jet power is rather modest. This does not come as a surprise since the magnetic field controls the
low-energy synchrotron output, which is limited for these types of FSRQs that are Compton dominated \citep{sahakyan20}.\\
In Figure \ref{fig:pvl} we show the total jet power $P_{jet}$ on the y-axis and the accretion disk luminosity $L_{disk}$ on the x-axis for the
different accretion efficiencies $\eta$. Their color coding is given in the legend. The dashed gray line shows the relation $P_{jet}$=$L_{disk}$.
According to the scaling relation of jets originating from both, stellar black holes and SMBHs \citep{nemmen12}, the power carried in the form of radiation
contributes to $\sim$10\% to the jet power thereby being approximately one order of magnitude less. This holds true also assuming one proton
per emitting electron as in our model \citep{ghisellini14}. In fact, if $P_{jet}$~$\ngeq$~$P_{rad}$, 
the jet power would be used up entirely to emitting radiation only. Thus, the jet would cease and the galaxy would not appear as a blazar.
Similarly the findings by \cite{celotti08} using a sizable number of blazars show that the total jet power is between 10 to 100 times
larger that the accretion disk luminosity. In Figure \ref{fig:pvl} only the total jet powers computed for low efficiencies $\eta$$<$0.15
result in $P_{jet}$ being approximately one order of magnitude larger than $L_{disk}$, thus suggesting the lower efficiencies are favored, thereby
supporting the best-fit model.\\

\begin{table*}[htbp]
\centering
\renewcommand{\arraystretch}{1.2}
\begin{tabular}{l|lllllll}
\toprule

\textbf{Parameter}            & \multicolumn{6}{c}{\textbf{Value}}                            & \textbf{Units}          \\ \midrule
$\eta$                  	& 0.057 & 0.083 & 0.100 & 0.150 & 0.200 & 0.300 & \nodata \\ \midrule
$N$               		& $11.50^{+1.29}_{-1.06}$ & $9.01^{+1.30}_{-1.14}$ & $7.55^{+1.27}_{-1.13}$ & $8.74^{+1.28}_{-1.56}$ & $7.98^{+1.26}_{-1.15}$ & $6.02^{+1.04}_{-0.92}$ & $10^{2}\text{ cm}^{-3}$ \\
$M_{BH}$          	& $1.19^{+0.23}_{-0.18}$  & $1.13^{+0.23}_{-0.17}$ & $1.27^{+0.29}_{-0.24}$ & $1.43^{+0.19}_{-0.19}$ & $1.32^{+0.29}_{-0.26}$ & $1.07^{+0.24}_{-0.19}$ & $10^{9} M_{\odot}$ \\
$R_{\text{diss}}$ 	& $1.23^{+0.31}_{-0.19}$  & $1.52^{+0.44}_{-0.28}$ & $1.81^{+0.49}_{-0.29}$ & $2.13^{+0.41}_{-0.31}$ & $2.45^{+0.68}_{-0.43}$ & $2.76^{+0.53}_{-0.41}$ & $10^{18}\text{ cm}$ \\
$B$               		& $7.89^{+0.12}_{-0.10}$  & $6.69^{+0.13}_{-0.09}$ & $4.13^{+0.74}_{-0.63}$ & $2.22^{+0.45}_{-0.28}$ & $1.94^{+0.35}_{-0.29}$ & $2.94^{+0.59}_{-0.50}$ & $10^{-2}\text{ G}$ \\
$L_{\text{disk}}$ 	& $1.00^{+0.12}_{-0.12}$  & $1.41^{+0.18}_{-0.18}$ & $2.01^{+0.25}_{-0.24}$ & $3.30^{+0.35}_{-0.32}$ & $4.22^{+0.56}_{-0.57}$ & $5.72^{+0.81}_{-0.63}$ & $10^{46}\text{ erg s}^{-1}$\\
$\Gamma$          	& $2.00^{+0.22}_{-0.24}$  & $1.89^{+0.24}_{-0.20}$ & $1.56^{+0.12}_{-0.10}$ & $1.33^{+0.12}_{-0.11}$ & $1.32^{+0.13}_{-0.10}$ & $1.48^{+0.15}_{-0.15}$ & $10^{1}$\\ \hline
\end{tabular}
\caption{The models were subjected to Markov chain Monte Carlo (MCMC) sampling. Specifically, the provided values represent the 16th, 50th, and 84th percentiles of samples within the marginalized distributions.}
\label{tab:MCMC}
\end{table*}

\begin{table*}[htbp]
\centering
\renewcommand{\arraystretch}{1.1}
\footnotesize
\begin{tabular}{l|ccccccc}
\toprule
\textbf{Parameter}            & \multicolumn{6}{c}{\textbf{Value}}                            & \textbf{Units}          \\ \midrule
$\eta$			& 0.057 & 0.083 & 0.100 & 0.150 & 0.200 & 0.300 & \nodata \\ \midrule
$P_{B}$			& 3.74 & 3.61 & 1.34 & 0.37 & 0.35 & 0.86 & $10^{43} \text{ erg s}^{-1}$ \\
$P_{e}$			& 7.37 & 7.85 & 7.71 & 9.49 & 10.60 & 11.89 & $10^{45} \text{ erg s}^{-1}$ \\
$P_{p}$			& 2.57 & 2.92 & 2.38 & 2.67 & 2.99 & 3.95 & $10^{46} \text{ erg s}^{-1}$\\
$P_{\text{rad}}$		& 9.53 & 8.08 & 5.98 & 5.08 & 4.9 & 5.24 & $10^{44} \text{ erg s}^{-1}$ \\
$P_{\text{jet}}$ 		& 3.40 & 3.79 & 3.21 & 3.68 & 4.10 & 5.19 & $10^{46} \text{ erg s}^{-1}$ \\
\bottomrule
\end{tabular}
\caption{Luminosity components carried by the jet for the radiative components, the electrons, the magnetic fields, and for the cold protons in the jet for our different models. For calculating $P_{p}$, we are considering a ratio of cold protons to relativistic electrons of 0.10.}
\label{tab:jetpower}
\end{table*}

\subsection{Growth of the SMBH in \target}
Measurements of accretion efficiencies and masses of SMBHs in the early Universe can be constraining for the evolution and origin of the seed
black holes \citep{inayoshi20}. These seed black holes had only little time in an Eddington-limited process (for which the maximum luminosity
at which the outward radiation pressure balances the inward gravitational pull) to accrete into a SMBH of
$M_{BH}$$\approx$1.1$\times$$10^{9} M_{\odot}$. In the scenario proposed by \cite{shapiro05}, in which the growth of the initial mass at time
$t_{0}$ of the seed black hole $M_{BH}(t_{0})$ occurs at a constant Eddington ratio, the evolution over time is given by an exponential growth following
the equation:  
\begin{equation}
    M_{BH}(t_{0})=M_{BH}(t)\exp{\left[-\eta_{\text{Edd}}\dfrac{1-\eta}{\eta}\dfrac{t-t_{0}}{\tau_{\text{acc}}}\right]}.
\label{eq:MBH_Evo}
\end{equation}
In this equation $\eta_{Edd}$ is the efficiency of the accretion luminosity given by $L_{acc}$/$L_{Edd}$ and $\tau_{acc}$ is the characteristic accretion
timescale. Specifically, the latter is defined as:
\begin{equation}
    \tau_{\text{acc}}\equiv\dfrac{M_{BH}c^{2}}{L_{\text{Edd}}}\approx0.45\mu^{-1}_{e} \text{ Gyr}
\end{equation}
which is independent of $M_{BH}$ and $\mu_{e}$ is the average molecular weight per electron given by
\begin{equation}
\mu_{e}=\dfrac{1}{1-Y/2}\approx\dfrac{8}{7}
\end{equation}
assuming the helium abundance $Y\approx0.25$ as provided by \cite{cyburt03}. 

\begin{figure}[htbp]
    \centering
    \includegraphics[width=0.47\textwidth]{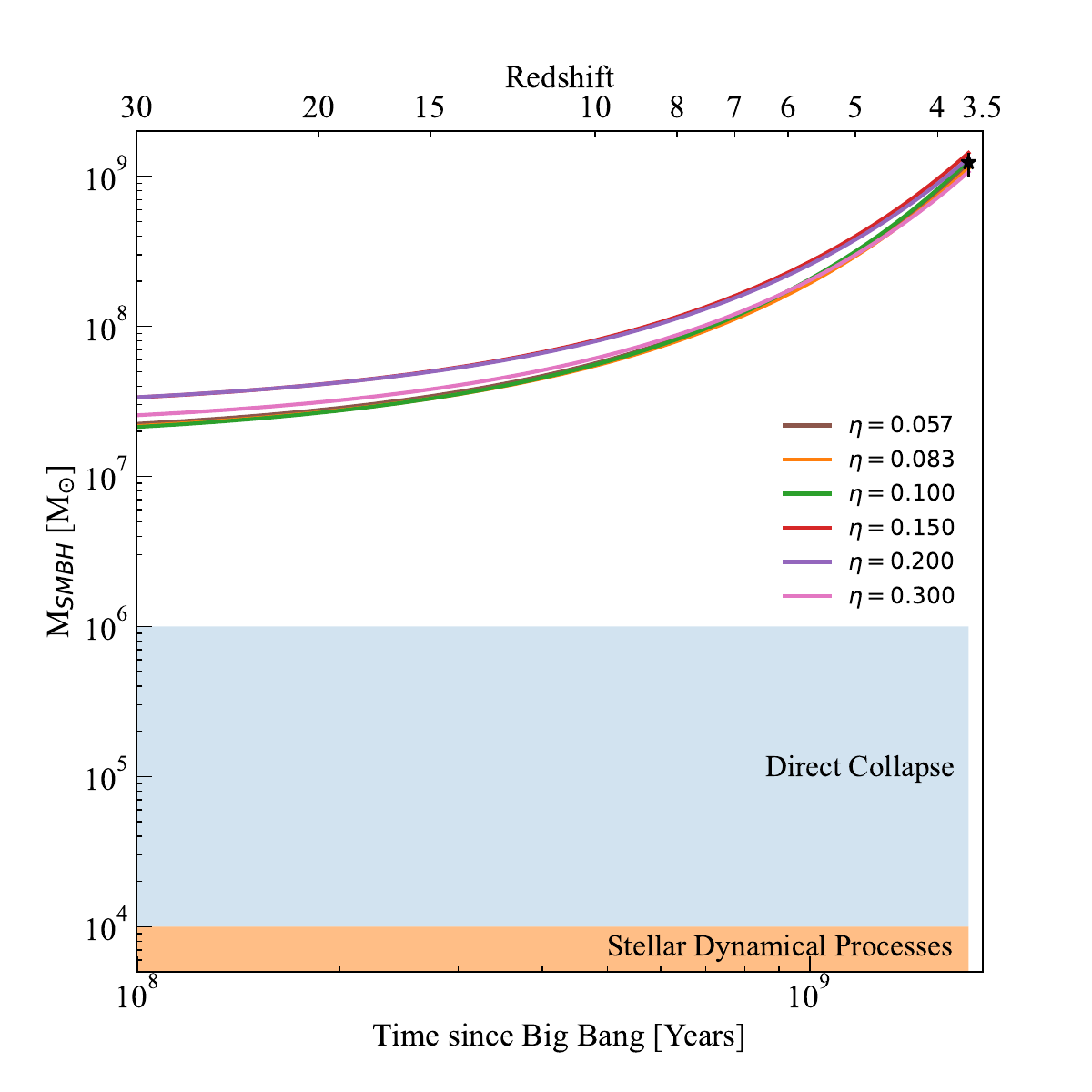}
    \caption{$M_{BH}$ evolution for \target~since redshift $z\approx30$ to $z\approx3.65$ (corresponding to $t=1.722$ Gyr). Colored lines represent the evolution for different values of accretion efficiency $\eta$.
     The shaded blue region correspond to the expected mass ranges direct collapse black holes ($M_{\text{seed}}\sim10^{4}-10^{6}M_{\odot}$). None of the different accretion efficiencies $\eta$ are able to
     trace back a SMBH of mass $M_{BH}$$\approx$10$^{9}$ $M_{\odot}$ supporting alternative evolution scenarios (see text for discussion).}
    \label{fig:BH_evo}
\end{figure}

The exponential growth model of the seed black holes is shown in Figure \ref{fig:BH_evo} for the different disk accretion efficiencies that
are represented by the different colors following the previous color coding. The error bars of the mass value at $z=3.65$ come from the
uncertainties of the different mass values computed for the different efficiencies. The shaded areas represent the predicted mass intervals of
black holes due to stellar dynamical processes ($M_{BH}\ge10^{3} - 10^{4} M_{\odot}$) and due to direct collapse ($M_{BH}\sim10^{4} - 10^{6} M_{\odot}$)
\citep{valiante16}. Direct collapse provides the most massive seed black holes. The formation of such heavy seeds faces specific issues.
However, very recent theoretical advancements through simulations of astroparticle physics including evaporating primordial black holes,
contracting dark matter halos, and Lyman-Werner photon production ease such tensions \citep{lu24a, lu24b}.
None of the different inferred efficiency values are able to trace back the evolution to $z\sim30$ when the first stars and galaxies are assumed to be in place.
This study shows that only efficiencies $\eta$$<$0.15 allow for jet powers that are significantly smaller than the accretion disk luminosity.
Such small efficiencies provide also the environment for a rapid black hole growth. Yet, high efficiencies caused by rapidly spinning black
holes \citep{thorne74} are needed to be able to form jets \citep{blandford77}, which is also supported by general relativistic
magnetohydrodynamic simulations \citep{mckinney06}. To reconcile the necessity for a rapidly spinning black hole and the inferred low
accretion efficiency, it is possible to envision that not all the energy of the infalling material is transformed into radiation (from heat),
but only a part is, while the rest goes into enhancing the magnetic field \citep{jolley08a}, which is necessary for the formation of the jet.
Thus, the total accretion efficiency is $\eta_{tot}=\eta_{r}+\eta_{j}$, where $\eta_{r}$ is the efficiency of converting the rest mass energy to
radiation and $\eta_{j}$ is the jet efficiency. For $\eta_{j}>0$, the jet can enhance angular momentum transport giving rise to higher
mass accretion rate. In other words, given an $\eta_{tot}$, $\eta_{r}$ can be lower favoring the accretion rate. Therefore, the presence
of a jet could in principle amplify the black hole growth rate \citep{blandford82, ghisellini13, ghisellini15}. This would allow a rapidly spinning black hole ($a=0.98$) to accrete into
a SMBH of $M_{BH}\approx10^{9} M_{\odot}$ by $z\approx6$ \citep{jolley08b}.\\
A further viable explanation for a rapid growth invokes episodes of super-Eddington accretion \citep[e.g.][]{banados21}. Recent, hydrodynamical
simulations of the black hole growth in the first billion years of the Universe show how frequent episodes of super-Eddington accretion
can lead to a significant black hole growth in a sophisticated arrangement of jet feedback \citep{massonneau23}. 
Other sophisticated simulations including mechanical feedback in early protogalaxies show that jets might regulate the BH growth \citep{su23, su24}.
Yet, such simulations rely on the simplifying assumption of the BH being fed within a single atomic halo, not accounting for magnetic fields and stellar physics.

\subsection{Conclusions}
This research describes the multifrequency campaign and its results of the blazar \target~at redshift $z=3.65$. This is one of the most
distant FSRQs detected at gamma-ray energies by the {\em Fermi} mission. In fact, its detection is facilitated by Compton dominance,
where the high-energy inverse-Compton component significantly outweighs the low-energy synchrotron component. This rather weak
synchrotron radiation and its peak, which is cosmologically shifted to lower frequencies, allow for the thermal emission by the accretion
disk to surface above the nonthermal synchrotron radiation.\\
The averaged SED is fit with a one-zone leptonic model. The best-fit result is obtained for a SMBH of $M_{BH}=1.1^{+0.2}_{-0.1}\times10^{9} M_{\odot}$.
The emitting region is outside the BLR but within the DT radius that contributes the most to the IC. The magnetic field of $B=6.56^{+0.13}_{-0.09}\times10^{-2}~G$ is
modest. This is expected as it controls the synchrotron output, which is limited due to the Compton dominance. As the inferred accretion efficiency is rather low
($\eta$=0.083), we also account for other accretion disk efficiencies, which result in different jet powers. Only efficiencies of $\eta$$<$0.15 yield total jet powers
that comply with the universal scaling law established by \cite{nemmen12} and produce accretion disk luminosities that align with the range identified in
population studies. Thus, the low efficiency is not solely inferred through the SED modeling and its physical implications discussed but also through the energetics.\\
The evolution study for the efficiencies using the model by \cite{shapiro05} cannot trace back such a massive SMBH to the time when the first stars and galaxies 
came into place ($z$$\approx$$30$). This holds true even for the most favorable physical condition for a rapid evolution due to a low efficiency. This result suggests
that accretion in an Eddington-limited process is not enough to grow the SMBH in \target~to $M_{BH}$$\approx$$1.1\times10^{9} M_{\odot}$. Either this mass
evolved from a much larger initial black hole mass or a more rapid evolution is needed. For a more rapid evolution, super-Eddington accretion must be invoked.
Alternatively, part of the gravitational energy causing accretion must be converted to mechanical energy (rather than to radiation only) to form the jet,
which allows for a more efficient accretion and thus a more rapid evolution.\\

\facilities{Fermi, XMM, Swift, WISE, GAIA}

\bibliographystyle{aasjournal}

\begin{thebibliography}{99}

\bibitem[Abdollahi et al.(2022)]{abdollahi22} Abdollahi, S., Acero, F., Baldini, L., et al.\ 2022, \apjs, 260, 53. doi:10.3847/1538-4365/ac6751

\bibitem[Ackermann et al.(2017)]{ackermann17} Ackermann, M., Ajello, M., Baldini, L., et al.\ 2017, \apjl, 837, L5. doi:10.3847/2041-8213/aa5fff

\bibitem[Ajello et al.(2022)]{Ajello22} Ajello, M., Baldini, L., Ballet, J., et al.\ 2022, \apjs, 263, 24. doi:10.3847/1538-4365/ac9523

\bibitem[Albareti et al.(2017)]{alberati17} Albareti, F.~D., Allende Prieto, C., Almeida, A., et al.\ 2017, \apjs, 233, 25. doi:10.3847/1538-4365/aa8992

\bibitem[Albareti et al.(2017)]{alberti17} Albareti, F.~D., Allende Prieto, C., Almeida, A., et al.\ 2017, \apjs, 233, 25. doi:10.3847/1538-4365/aa8992

\bibitem[Atwood et al.(2009)]{atwood09} Atwood, W.~B., Abdo, A.~A., Ackermann, M., et al.\ 2009, \apj, 697, 1071. doi:10.1088/0004-637X/697/2/1071

\bibitem[B{\l}a{\.z}ejowski et al.(2000)]{bazejowski00} B{\l}a{\.z}ejowski, M., Sikora, M., Moderski, R., et al.\ 2000, \apj, 545, 107. doi:10.1086/317791

\bibitem[Ba{\~n}ados et al.(2021)]{banados21} Ba{\~n}ados, E., Mazzucchelli, C., Momjian, E., et al.\ 2021, \apj, 909, 80. doi:10.3847/1538-4357/abe239

\bibitem[Ballet et al.(2023)]{ballet23} Ballet, J., Bruel, P., Burnett, T.~H., et al.\ 2023, arXiv:2307.12546. doi:10.48550/arXiv.2307.12546

\bibitem[Bambi et al.(2021)]{bambi21} Bambi, C., Brenneman, L.~W., Dauser, T., et al.\ 2021, \ssr, 217, 65. doi:10.1007/s11214-021-00841-8 

\bibitem[Bardeen(1973)]{bardeen73} Bardeen, J.~M.\ 1973, Black Holes (Les Astres Occlus), 215

\bibitem[Becker et al.(1991)]{becker91} Becker, R.~H., White, R.~L., \& Edwards, A.~L.\ 1991, \apjs, 75, 1. doi:10.1086/191529

\bibitem[Becker et al.(1995)]{becker95} Becker, R.~H., White, R.~L., \& Helfand, D.~J.\ 1995, \apj, 450, 559. doi:10.1086/176166

\bibitem[Blandford \& Rees(1974)]{blandford74} Blandford, R.~D. \& Rees, M.~J.\ 1974, \mnras, 169, 395. doi:10.1093/mnras/169.3.395

\bibitem[Blandford \& Znajek(1977)]{blandford77} Blandford, R.~D. \& Znajek, R.~L.\ 1977, \mnras, 179, 433. doi:10.1093/mnras/179.3.433

\bibitem[Blandford \& Payne(1982)]{blandford82} Blandford, R.~D. \& Payne, D.~G.\ 1982, \mnras, 199, 883. doi:10.1093/mnras/199.4.883

\bibitem[Bloom \& Marscher(1996)]{bloom96} Bloom, S.~D. \& Marscher, A.~P.\ 1996, \apj, 461, 657. doi:10.1086/177092

\bibitem[Bottacini et al.(2010)]{bottacini10} Bottacini, E., Ajello, M., Greiner, J., et al.\ 2010, \aap, 509, A69. doi:10.1051/0004-6361/200913260

\bibitem[Cardelli et al.(1989)]{cardelli89} Cardelli, J.~A., Clayton, G.~C., \& Mathis, J.~S.\ 1989, \apj, 345, 245. doi:10.1086/167900

\bibitem[Celotti \& Ghisellini(2008)]{celotti08} Celotti, A. \& Ghisellini, G.\ 2008, \mnras, 385, 283. doi:10.1111/j.1365-2966.2007.12758.x

\bibitem[Cleary et al.(2007)]{cleary07} Cleary, K., Lawrence, C.~R., Marshall, J.~A., et al.\ 2007, \apj, 660, 117. doi:10.1086/511969

\bibitem[Condon et al.(1998)]{condon98} Condon, J.~J., Cotton, W.~D., Greisen, E.~W., et al.\ 1998, \aj, 115, 1693. doi:10.1086/300337

\bibitem[Cyburt et al.(2003)]{cyburt03} Cyburt, R.~H., Fields, B.~D., \& Olive, K.~A.\ 2003, Physics Letters B, 567, 227. doi:10.1016/j.physletb.2003.06.026

\bibitem[Dermer et al.(1992)]{dermer92} Dermer, C.~D., Schlickeiser, R., \& Mastichiadis, A.\ 1992, \aap, 256, L27

\bibitem[Finke(2016)]{finke16} Finke, J.\ 2016, arXiv:1602.05965. doi:10.48550/arXiv.1602.05965

\bibitem[Fossati et al.(1998)]{fossati98} Fossati, G., Maraschi, L., Celotti, A., et al.\ 1998, \mnras, 299, 433. doi:10.1046/j.1365-8711.1998.01828.x

\bibitem[Franceschini et al.(2008)]{franceschini08} Franceschini, A., Rodighiero, G., \& Vaccari, M.\ 2008, \aap, 487, 837. doi:10.1051/0004-6361:200809691

\bibitem[Gaia Collaboration et al.(2016)]{prusti16} Gaia Collaboration, Prusti, T., de Bruijne, J.~H.~J., et al.\ 2016, \aap, 595, A1. doi:10.1051/0004-6361/201629272

\bibitem[Gaia Collaboration et al.(2018)]{brown18} Gaia Collaboration, Brown, A.~G.~A., Vallenari, A., et al.\ 2018, \aap, 616, A1. doi:10.1051/0004-6361/201833051

\bibitem[Gaia Collaboration et al.(2023)]{vallenari23} Gaia Collaboration, Vallenari, A., Brown, A.~G.~A., et al.\ 2023, \aap, 674, A1. doi:10.1051/0004-6361/202243940

\bibitem[Georganopoulos \& Kazanas(2003)]{georganopoulos03} Georganopoulos, M. \& Kazanas, D.\ 2003, \apjl, 594, L27. doi:10.1086/378557

\bibitem[Ghisellini \& Madau(1996)]{ghisellini96} Ghisellini, G. \& Madau, P.\ 1996, \mnras, 280, 67. doi:10.1093/mnras/280.1.67

\bibitem[Ghisellini \& Tavecchio(2009)]{ghisellini09} Ghisellini, G. \& Tavecchio, F.\ 2009, \mnras, 397, 985. doi:10.1111/j.1365-2966.2009.15007.x

\bibitem[Ghisellini et al.(2010)]{ghisellini10} Ghisellini, G., Della Ceca, R., Volonteri, M., et al.\ 2010, \mnras, 405, 387. doi:10.1111/j.1365-2966.2010.16449.x

\bibitem[Ghisellini et al.(2013)]{ghisellini13} Ghisellini, G., Haardt, F., Della Ceca, R., et al.\ 2013, \mnras, 432, 2818. doi:10.1093/mnras/stt637

\bibitem[Ghisellini et al.(2014)]{ghisellini14} Ghisellini, G., Tavecchio, F., Maraschi, L., et al.\ 2014, \nat, 515, 376. doi:10.1038/nature13856

\bibitem[Ghisellini(2015)]{ghisellini15} Ghisellini, G.\ 2015, Journal of High Energy Astrophysics, 7, 163. doi:10.1016/j.jheap.2015.03.002

\bibitem[Gregory \& Condon(1991)]{gregory91} Gregory, P.~C. \& Condon, J.~J.\ 1991, \apjs, 75, 1011. doi:10.1086/191559

\bibitem[Griffith et al.(1990)]{griffith90} Griffith, M., Langston, G., Heflin, M., et al.\ 1990, \apjs, 74, 129. doi:10.1086/191495

\bibitem[Helmboldt et al.(2007)]{helmboldt07} Helmboldt, J.~F., Taylor, G.~B., Tremblay, S., et al.\ 2007, \apj, 658, 203. doi:10.1086/511005

\bibitem[Inayoshi et al.(2020)]{inayoshi20} Inayoshi, K., Visbal, E., \& Haiman, Z.\ 2020, \araa, 58, 27. doi:10.1146/annurev-astro-120419-014455

\bibitem[Jansen et al.(2001)]{jansen01} Jansen, F., Lumb, D., Altieri, B., et al.\ 2001, \aap, 365, L1. doi:10.1051/0004-6361:20000036

\bibitem[Jarrett et al.(2011)]{jarrett11} Jarrett, T.~H., Cohen, M., Masci, F., et al.\ 2011, \apj, 735, 112. doi:10.1088/0004-637X/735/2/112

\bibitem[Jolley \& Kuncic(2008)]{jolley08a} Jolley, E.~J.~D. \& Kuncic, Z.\ 2008, \apj, 676, 351. doi:10.1086/527312

\bibitem[Jolley \& Kuncic(2008)]{jolley08b} Jolley, E.~J.~D. \& Kuncic, Z.\ 2008, \mnras, 386, 989. doi:10.1111/j.1365-2966.2008.13082.x

\bibitem[Kalberla et al.(2005)]{kalberla05} Kalberla, P.~M.~W., Burton, W.~B., Hartmann, D., et al.\ 2005, \aap, 440, 775. doi:10.1051/0004-6361:20041864

\bibitem[Kaspi et al.(2005)]{kaspi05} Kaspi, S., Maoz, D., Netzer, H., et al.\ 2005, \apj, 629, 61. doi:10.1086/431275

\bibitem[Kaspi et al.(2007)]{kaspi07} Kaspi, S., Brandt, W.~N., Maoz, D., et al.\ 2007, \apj, 659, 997. doi:10.1086/512094

\bibitem[Lu et al.(2024a)]{lu24a} Lu, Y., Picker, Z.~S.~C., \& Kusenko, A.\ 2024a, \prl, 133, 091001. doi:10.1103/PhysRevLett.133.091001

\bibitem[Lu et al.(2024b)]{lu24b} Lu, Y., Picker, Z.~S.~C., \& Kusenko, A.\ 2024b, \prd, 109, 123016. doi:10.1103/PhysRevD.109.123016

\bibitem[Maraschi \& Tavecchio(2003)]{maraschi03} Maraschi, L. \& Tavecchio, F.\ 2003, \apj, 593, 667. doi:10.1086/342118

\bibitem[Marscher \& Gear(1985)]{marscher85} Marscher, A.~P. \& Gear, W.~K.\ 1985, \apj, 298, 114. doi:10.1086/163592

\bibitem[Massaro \& D'Abrusco(2016)]{massaro16} Massaro, F. \& D'Abrusco, R.\ 2016, \apj, 827, 67. doi:10.3847/0004-637X/827/1/67

\bibitem[Massaro et al.(2011)]{massaro11} Massaro, F., D'Abrusco, R., Ajello, M., et al.\ 2011, \apjl, 740, L48. doi:10.1088/2041-8205/740/2/L48

\bibitem[Massonneau et al.(2023)]{massonneau23} Massonneau, W., Volonteri, M., Dubois, Y., et al.\ 2023, \aap, 670, A180. doi:10.1051/0004-6361/202243170

\bibitem[Matsuoka et al.(2011)]{matsuoka11} Matsuoka, Y., Ienaka, N., Kawara, K., et al.\ 2011, \apj, 736, 119. doi:10.1088/0004-637X/736/2/119

\bibitem[McKinney(2006)]{mckinney06} McKinney, J.~C.\ 2006, \mnras, 368, 1561. doi:10.1111/j.1365-2966.2006.10256.x

\bibitem[Natarajan(2014)]{natarajan14} Natarajan, P.\ 2014, General Relativity and Gravitation, 46, 1702. doi:10.1007/s10714-014-1702-6

\bibitem[Nemmen et al.(2012)]{nemmen12} Nemmen, R.~S., Georganopoulos, M., Guiriec, S., et al.\ 2012, Science, 338, 1445. doi:10.1126/science.1227416

\bibitem[Urry \& Padovani(1995)]{urry95} Urry, C.~M. \& Padovani, P.\ 1995, \pasp, 107, 803. doi:10.1086/133630

\bibitem[Padovani \& Giommi(1995)]{padovani95} Padovani, P. \& Giommi, P.\ 1995, \apj, 444, 567. doi:10.1086/175631

\bibitem[Padovani(2016)]{padovani16} Padovani, P.\ 2016, \aapr, 24, 13. doi:10.1007/s00159-016-0098-6

\bibitem[Petrov \& Taylor(2011)]{petrov11} Petrov, L. \& Taylor, G.~B.\ 2011, \aj, 142, 89. doi:10.1088/0004-6256/142/3/89

\bibitem[Plavin et al.(2019)]{plavin19} Plavin, A.~V., Kovalev, Y.~Y., \& Petrov, L.~Y.\ 2019, \apj, 871, 143. doi:10.3847/1538-4357/aaf650

\bibitem[Plotkin et al.(2012)]{plotkin12} Plotkin, R.~M., Anderson, S.~F., Brandt, W.~N., et al.\ 2012, \apjl, 745, L27. doi:10.1088/2041-8205/745/2/L27

\bibitem[Poole et al.(2008)]{poole08} Poole, T.~S., Breeveld, A.~A., Page, M.~J., et al.\ 2008, \mnras, 383, 627. doi:10.1111/j.1365-2966.2007.12563.x

\bibitem[Pushkarev et al.(2010)]{pushkarev10} Pushkarev, A.~B., Kovalev, Y.~Y., \& Lister, M.~L.\ 2010, \apjl, 722, L7. doi:10.1088/2041-8205/722/1/L7

\bibitem[Rau et al.(2012)]{rau12} Rau, A., Schady, P., Greiner, J., et al.\ 2012, \aap, 538, A26. doi:10.1051/0004-6361/201118159

\bibitem[Rees(1971)]{rees71} Rees, M.~J.\ 1971, \nat, 229, 312. doi:10.1038/229312a0

\bibitem[Rees(1978)]{rees78} Rees, M.~J.\ 1978, The Observatory, 98, 210

\bibitem[Reynolds(2019)]{reynolds19} Reynolds, C.~S.\ 2019, Nature Astronomy, 3, 41. doi:10.1038/s41550-018-0665-z

\bibitem[Roming et al.(2005)]{roming05} Roming, P.~W.~A., Kennedy, T.~E., Mason, K.~O., et al.\ 2005, \ssr, 120, 95. doi:10.1007/s11214-005-5095-4

\bibitem[Rueda-Becerril et al.(2021)]{rueda21} Rueda-Becerril, J.~M., Harrison, A.~O., \& Giannios, D.\ 2021, \mnras, 501, 4092. doi:10.1093/mnras/staa3925

\bibitem[Sahakyan et al.(2020)]{sahakyan20} Sahakyan, N., Israyelyan, D., Harutyunyan, G., et al.\ 2020, \mnras, 498, 2594. doi:10.1093/mnras/staa2477

\bibitem[Sambruna(1997)]{sambruna97} Sambruna, R.~M.\ 1997, \apj, 487, 536. doi:10.1086/304640

\bibitem[Sambruna et al.(2006)]{sambruna06} Sambruna, R.~M., Gliozzi, M., Tavecchio, F., et al.\ 2006, \apj, 652, 146. doi:10.1086/507420

\bibitem[Sawicki(2012)]{sawicki12} Sawicki, M.\ 2012, \pasp, 124, 1208. doi:10.1086/668636

\bibitem[Scheuer(1974)]{scheuer74} Scheuer, P.~A.~G.\ 1974, \mnras, 166, 513. doi:10.1093/mnras/166.3.513

\bibitem[Schlegel et al.(1998)]{schlegel98} Schlegel, D.~J., Finkbeiner, D.~P., \& Davis, M.\ 1998, \apj, 500, 525. doi:10.1086/305772

\bibitem[Shapiro(2005)]{shapiro05} Shapiro, S.~L.\ 2005, \apj, 620, 59. doi:10.1086/427065

\bibitem[Sikora et al.(1994)]{sikora94} Sikora, M., Begelman, M.~C., \& Rees, M.~J.\ 1994, \apj, 421, 153. doi:10.1086/173633

\bibitem[Stecker et al.(1992)]{stecker92} Stecker, F.~W., de Jager, O.~C., \& Salamon, M.~H.\ 1992, \apjl, 390, L49. doi:10.1086/186369

\bibitem[Su et al.(2023)]{su23} Su, K.-Y., Bryan, G.~L., Haiman, Z., et al.\ 2023, \mnras, 520, 4258. doi:10.1093/mnras/stad252

\bibitem[Su et al.(2024)]{su24} Su, K.-Y., Bryan, G., \& Haiman, Z.\ 2024, arXiv:2409.12250. doi:10.48550/arXiv.2409.12250

\bibitem[Tanaka \& Haiman(2009)]{tanaka09} Tanaka, T. \& Haiman, Z.\ 2009, \apj, 696, 1798. doi:10.1088/0004-637X/696/2/1798

\bibitem[Thorne(1974)]{thorne74} Thorne, K.~S.\ 1974, \apj, 191, 507. doi:10.1086/152991

\bibitem[Tramacere et al.(2009)]{tramacere09} Tramacere, A., Giommi, P., Perri, M., et al.\ 2009, \aap, 501, 879. doi:10.1051/0004-6361/200810865

\bibitem[Tramacere et al.(2011)]{tramacere11} Tramacere, A., Massaro, E., \& Taylor, A.~M.\ 2011, \apj, 739, 66. doi:10.1088/0004-637X/739/2/66

\bibitem[Tramacere(2020)]{tramacere20} Tramacere, A.\ 2020, Astrophysics Source Code Library. ascl:2009.001

\bibitem[Tramacere et al.(2022)]{tramacere22} Tramacere, A., Sliusar, V., Walter, R., et al.\ 2022, \aap, 658, A173. doi:10.1051/0004-6361/202142003

\bibitem[Valiante et al.(2016)]{valiante16} Valiante, R., Schneider, R., Volonteri, M., et al.\ 2016, \mnras, 457, 3356. doi:10.1093/mnras/stw225

\bibitem[Volonteri et al.(2008)]{volonteri08} Volonteri, M., Lodato, G., \& Natarajan, P.\ 2008, \mnras, 383, 1079. doi:10.1111/j.1365-2966.2007.12589.x

\bibitem[Volonteri et al.(2011)]{volonteri11} Volonteri, M., Haardt, F., Ghisellini, G., et al.\ 2011, \mnras, 416, 216. doi:10.1111/j.1365-2966.2011.19024.x

\bibitem[Weymann et al.(1981)]{weymann81} Weymann, R.~J., Carswell, R.~F., \& Smith, M.~G.\ 1981, \araa, 19, 41. doi:10.1146/annurev.aa.19.090181.000353

\bibitem[White \& Becker(1992)]{white92} White, R.~L. \& Becker, R.~H.\ 1992, \apjs, 79, 331. doi:10.1086/191656

\bibitem[Wright et al.(2010)]{wright10} Wright, E.~L., Eisenhardt, P.~R.~M., Mainzer, A.~K., et al.\ 2010, \aj, 140, 1868. doi:10.1088/0004-6256/140/6/1868

\bibitem[Wright et al.(2010)]{wright10} Wright, E.~L., Eisenhardt, P.~R.~M., Mainzer, A.~K., et al.\ 2010, \aj, 140, 1868. doi:10.1088/0004-6256/140/6/1868

\bibitem[Yoo \& Miralda-Escud{\'e}(2004)]{yoo04} Yoo, J. \& Miralda-Escud{\'e}, J.\ 2004, \apjl, 614, L25. doi:10.1086/425416

\bibitem[Zeng et al.(2021)]{zeng21} Zeng, H., Petrosian, V., \& Yi, T.\ 2021, \apj, 913, 120. doi:10.3847/1538-4357/abf65e

\end{thebibliography}

\end{document}